\documentclass{article}
\usepackage[utf8]{inputenc}
\usepackage{graphicx}
\usepackage{float}
\usepackage{authblk}

\title{Automated Superconducting Qubit Characterisation Platform Based on a Modified 3D Printer}
\author[1]{Haochen Li}
\author[2]{Soe Gon Yee Thant}
\author[3]{Rainer Dumke}
\affil[1,2,3]{Department of Physical and Mathematical Sciences, Nanyang Technological University}
\affil[3]{Centre for Quantum Technologies, National University of Singapore}
\date{}

\begin{document}

\maketitle

\section*{Abstract}

Josephson Junctions are important components in superconducting qubits. It introduces anharmonicity to the energy level spacings of the qubit which allow us to identify two unique quantum energy states for computing. It is difficult to fabricate multiple junctions within the same desired parameter range. Characterisation of the junctions is, therefore, a necessary step after fabrication. In particular, the critical current of the junctions is determined by measuring their normal state resistance. This is done via two-point or four-point resistance measurement at a manual probe station which is a time-consuming process, especially for wafer-scale fabrication. This bottleneck can be circumvented by automation with object detection. The base of the automated probe station is a 3D printer modified with multiple Arduino Uno microcontrollers and motorised linear stages. The automation process is achieved via auto-alignment of the probes and an automatic measurement procedure. As a result, the fully automated process will take about 27-29 seconds to measure the resistance of one junction which saves 28-51\% of the time compared to the manual probe station and can be unsupervised. Due to the reuse of a commercial 3D printer, the cost of this system is 800 SGD which is much less than comparable commercial solutions.  

\newpage

\section{Introduction}

Josephson Junction (JJ) is a barrier that separates two superconductors by a thin insulating layer \cite{coupledsuperconductors, Josephsoneffect}. It acts as a non-linear inductor shunted with a capacitor in a superconducting circuit. This forms an anharmonic oscillator, and thus, non-linear energy spacing in the qubit. Through this non-linearity, a two-state system can be obtained which forms the backbone of a transmon qubit \cite{quantumguide}. In a multi-transmon system, it is crucial to have precise control over the qubit frequency. Therefore, the variations in the junction area and tunnel barrier thickness should be minimised as much as possible to optimize the system performance and prevent spectral overlap between circuits \cite{JJresistanceVariation}.
However, the fabrication process of the JJ involves numerous complex steps of aluminum layer deposition and oxidation which is hard to control precisely \cite{MetrologyOfQuantumControl}. This makes the fabrication difficult and susceptible to errors as there are many parameters required to be carefully controlled such as deposition angles, oxidation pressure, and oxidation time \cite{Soe}. This imperfect fabrication causes a large variation in the normal state resistance of the JJ, and the characterisation of the resistance is strongly required. The normal state resistance is important because the critical current of the qubits can be determined from the resistance according to the Ambegaokar-Baratoff relation.

\begin{equation}
    I_cR = \frac{\pi}{2e} \Delta
\end{equation}  
where $\Delta$ is the superconducting gap in Joules, $I_c$ is the critical current and R is the normal state resistance \cite{simplified_JJ_Fabrication}. From this critical current, the Josephson energy $E_J$ and working frequency range can thus be determined.

As the normal state resistance characterisation process following qubit fabrication is crucial, optimizing the measurement process is essential.
This paper presents an extension of a manual probe station previously designed for resistance measurement \cite{Soe}. The manual probe station uses linear stages to manually move probes in the X, Y, and Z axes which is time-consuming. This paper describes a design of an automated resistance measurement process by modifying a 3D printer with multiple Arduino Uno microcontrollers and motorised linear stages. All the components of the system are affordable and can be easily acquired. The approximate estimated cost of this system is around 800 SGD which costs much less than a comparable commercial probe station.

The automation is done by auto-alignment and auto-lowering of probes to JJ pads. When the probes touch the JJ pads, a resistance-measuring program is then launched. The safe measure of the system includes the immediate stop of the probe lowering triggered by the excessive force exerted by the probes on the qubit and the limitation of the maximum values of voltage and current to prevent damaging the JJ during the measurement \cite{Soe}.

\begin{figure}[H]
    \centering
    \includegraphics[width=\textwidth]{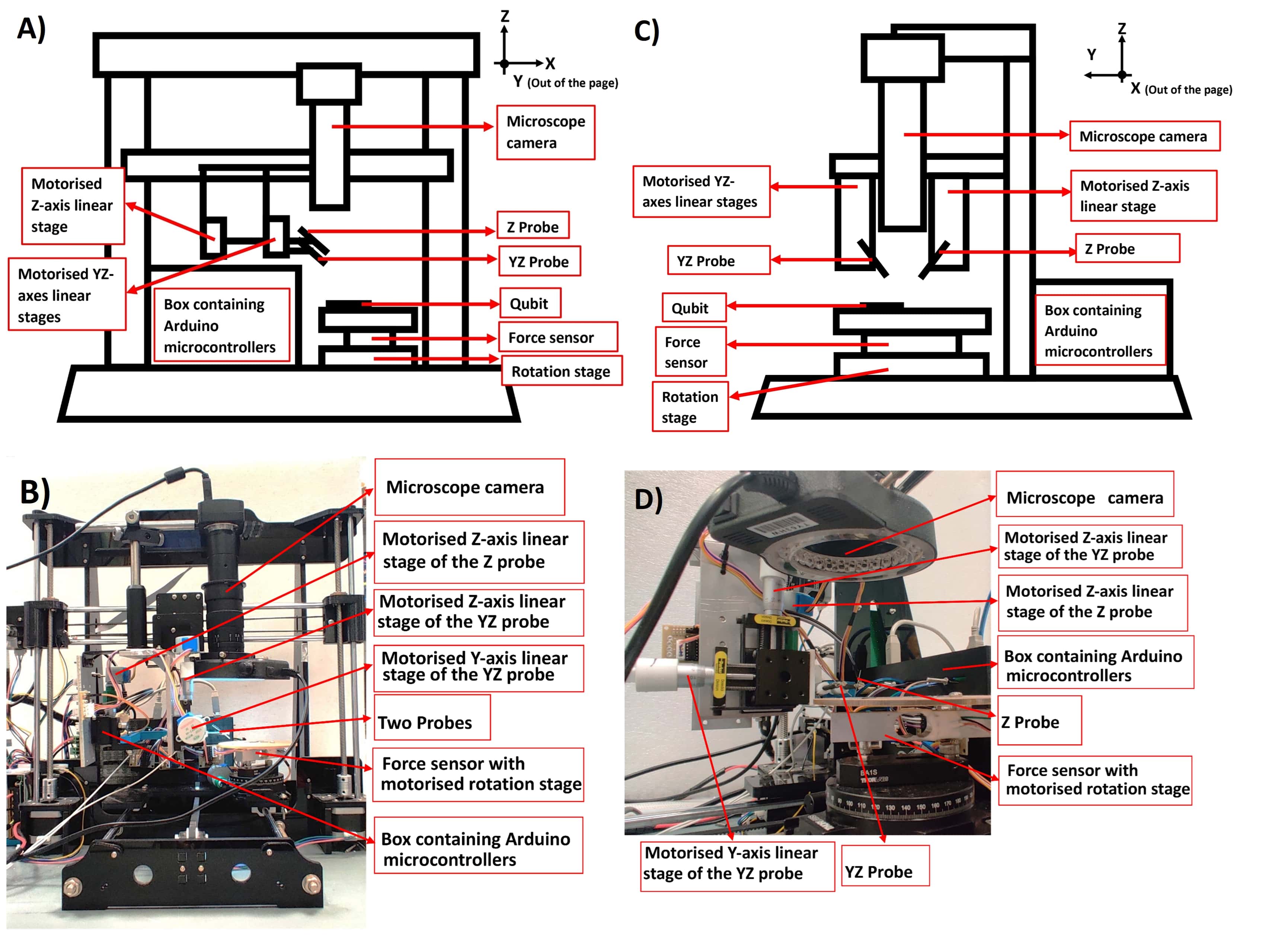}
    \caption{Photos B) and D) show different views of the characterisation platform. The two schematics A) and C) highlight important components of the system. A) and B) are the front views of the system while C) and D) are the side views of the system. The extruder from the 3D printer is replaced by multiple metal plates which hold motorised linear stages and a microscope camera. The probes are attached to the linear stages by 3D-printed adaptors. One probe is attached to a single Z-axis linear stage while the other probe is attached to Y and Z-axes linear stages. The heating bed of the 3D printer is removed to place a force sensor with motorised rotation stage. This is a platform for mounting the qubit chips and supervise the force exerted by the probes onto the qubit surface. The rotation stage can change the angle between the qubit and the probes so that they can be aligned.}
    \label{system photo}
\end{figure}

\section{Design of the System}
\subsection{Hardware Design}
The 3D printer we are using is TRONXY P802MA. Other models are equally viable for building this system. The position of its extruder is controlled by stepper motors. The extruder is secured on two horizontal sliding rods. It is also attached to a stepper motor via a timing belt. The rotation of the motor can slide the extruder horizontally to change its X-coordinate. The horizontal sliding rods are attached to two  Z-axis screw rods at both ends. The two Z-axis stepper motors can change the extruder’s Z-coordinate by lifting the horizontal rods up and down. The actual Y-coordinate of the extruder cannot be changed but its Y-coordinate relative to the heating bed can be changed by rotation of the stepper motor attached to the heating bed. The heating bed can thus move towards or away from the extruder to change the extruder’s relative Y-coordinate. The modified 3D printer is shown in Fig \ref{system photo}.

As shown in Fig \ref{system photo}, this characterisation platform is built based on the 3D printer. The extruder of the 3D printer is replaced by multiple metal plates. The plates hold one microscope camera and multiple motorised linear stages. The camera points directly down to the probes from above to have a clear view of the probes and the qubit in the two-dimensional XY plane. Two tungsten probes are attached to the linear stages with 3D-printed adaptors.
In this two-point resistance measurement, each probe is placed on a different side of the JJ. The first probe is attached to only one Z-axis linear stage to control its height above the qubit surface, and thus, it is called the Z probe. The second probe is attached to Y and Z axes linear stages to control both its height above the qubit surface and its horizontal distance to the Z probe, and thus, it is called the YZ probe. Every linear stage’s rotation knob is attached to a 28BYJ-48 electric stepper motor via a 3D-printed adaptor. The motors are controlled via the Arduino Uno microcontrollers. 

The reason for using the two-point measurement is that it is much easier to align the probes with the JJ compared to the four-point measurement. Additionally, since the JJ normal state resistance is usually in the range of kOhm while the typical contact resistance of a clean tungsten probe is only about 250 mOhm \cite{2pointResistance}, the contact resistance of the probes involved in the two-probes measurement is negligible. Thus, the two-point resistance measurement is accurate enough.

Moreover, the heating bed of the 3D printer is removed to place a force sensor. The force sensor consists of a strain gauge load cell and an HX711 load cell amplifier. A plastic circular platform is secured to the top of the load cell to mount qubits onto it. This force sensor is used to supervise the force exerted by the probes on the chip so that the probes exert enough force to scratch the oxidized aluminum $Al_2O_3$ layer without damaging the JJ. This is to make good physical contact between the probes and the JJ pads so that the measurement of the resistance is accurate.

\subsection{Control Design}

\begin{figure}[h]
    \centering
    \includegraphics[scale=0.5]{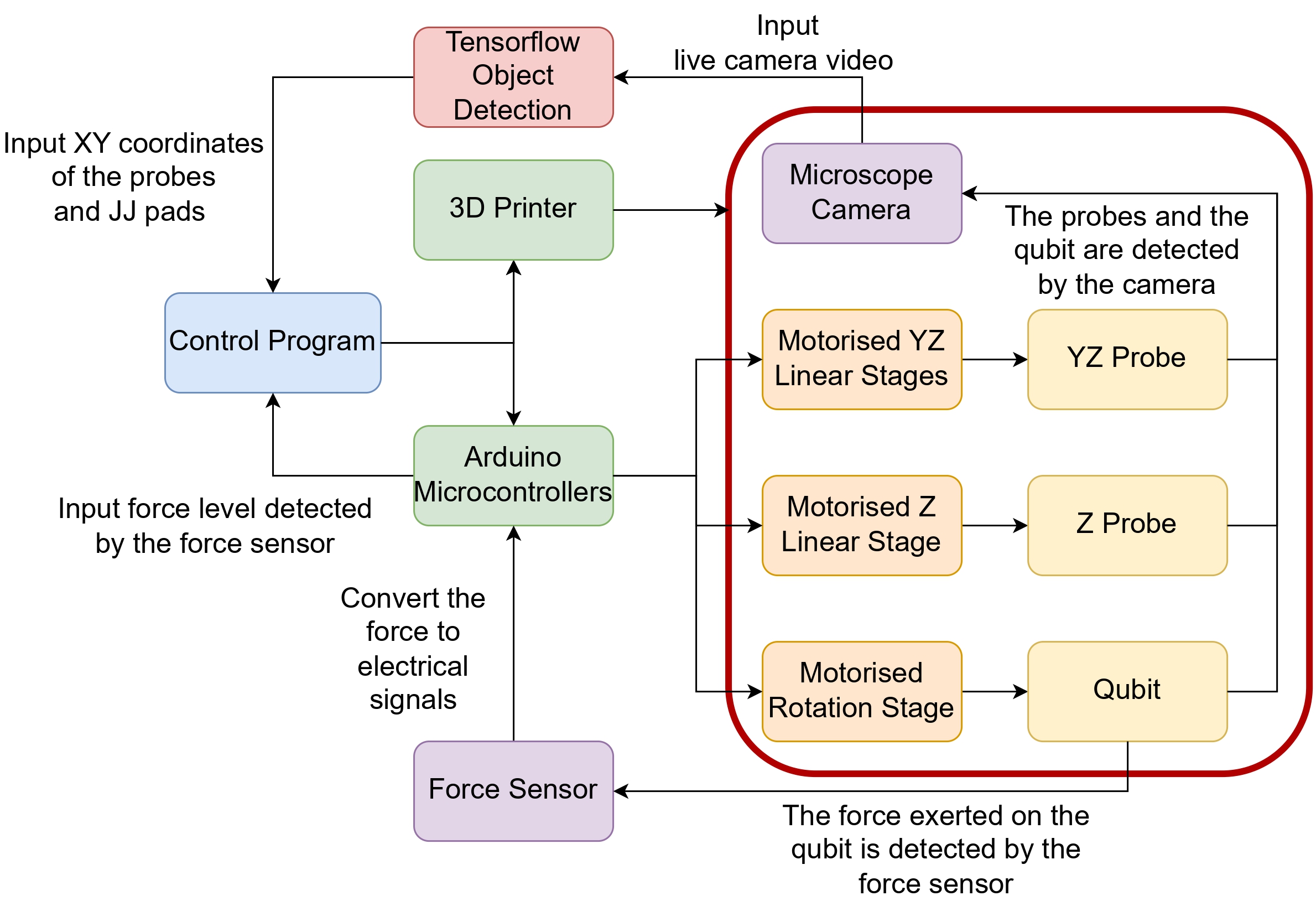}
    \caption{The block diagram shows the working mechanism of the system. For the alignment, a Tensorflow object detection program detects the objects from the live video sent from the microscope camera. This object detection program then extracts and inputs the XY coordinates of the probes’ tips and the JJ pads of the qubit into the control program. For the lowering, the control program takes the output from the force sensor which supervises the force exerted by the probes on the qubit. With all these data inputs, the control program can send commands to either the 3D printer to do the auto-alignment or the Arduino microcontrollers to do the auto-lowering. These two automated processes will be explained in detail in section 3.}
    \label{fig:block diagram}
\end{figure}

As shown in Fig \ref{fig:block diagram}, the program control can send commands to the 3D printer and Arduino microcontrollers to move the microscope camera and probes in the desired way. Via the Arduino microcontrollers, the program can adjust the positions of the probes. For the Z probe, only the Z-coordinate can be varied while for the YZ probe, both Y and Z coordinates can be varied. Their Z-coordinates are controlled to lower the probes to the qubit surface for the measurement. The Y-coordinate of the YZ probes is controlled so that the distance between the probes is smaller than the length of the JJ pads so that the probes can be positioned onto the pads. Additionally, the rotation stage below the mounting platform of the qubit is used to change the angle between the probes and the JJ pads to make sure that the probes can be placed on each side of the JJ. The force sensor attached to the mounting platform supervises the force exerted by the probes. This force level is inputted into the control program to achieve the auto-lowering of the probes to the qubit surface. The auto-lowering process will be explained in detail in section 3.2.

Since the 3D printer was designed to move the whole extruder and the heating bed, the commands sent to the 3D printer are used to move the microscope camera, probes, and the qubit. The microscope camera has a clear view of the probes and the qubit directly from above. The live video from the microscope camera can thus be used to identify the positions of the probes and JJ pads in the XY plane via an object detection program.

\begin{figure}
    \centering
    \includegraphics[scale=0.7]{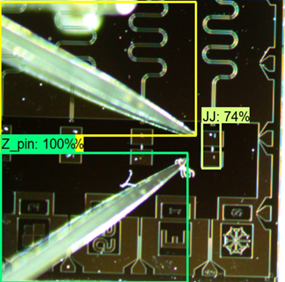}
    \caption{This figure shows the camera view window with the object detection label boxes. There are three types of objects, be they the YZ probe in the yellow box, the Z probe in the dark green box, and the JJ in the light green box. The label of the YZ probe is blocked by the label of the Z probe. The label box indicates the name of the detected object and the probability of successful detection. Since the camera is pointing down the Z-axis, the camera view shows the XY plane of the mounting platform. X-axis is horizontal while the Y-axis is vertical in this camera view. Each side of every label box can thus be identified as its X or Y coordinate. With these coordinates, the auto-alignment can be achieved which will be explained in detail in section 3.1.}
    \label{fig:tfod}
\end{figure}

The object detection program uses the Tensorflow model and OpenCV to detect objects shown on the live video. The model used is SSD Mobilenet V2 FPNLite 320x320. It is trained on COCO 2017 dataset and the training images are compressed to 320x320. It runs each frame in 22 ms and it has a mean average precision of 0.222  \cite{tensorflowmodelgarden2020}. The good balance between the speed and precision of this model is desired for the auto-alignment of the system discussed in section 3.1. Video capture from the OpenCV is used to capture the continuous live streams of the images from the microscope camera. This model takes an image input and returns boundary boxes of each detected object, the probabilities of successful detection of the objects, and the names of the detected objects as shown in Fig  \ref{fig:tfod} \cite{tensorflowmodelgarden2020}.

Fig \ref{fig:tfod} is a screenshot of the Tensorflow object detection window for the auto-alignment process. There are three boxes shown in the screenshot. Each box is labeled with its the identified object. The YZ probe is in the yellow box, the Z probe is in the dark green box, and the JJ is in the light green box. The probability number beside the name of the box is the probability of the successful recognition of the object.

Nevertheless, there are limitations of the object detection model to consider. The model training requires training images that depict the object in different backgrounds, lighting conditions, angles, and arrangements. The objects in the images are then labeled manually, and thus, it is usually time-consuming to train a model with numerous training images to obtain high precision. However, with the properly trained model, the auto-alignment usually only takes about 1 second to complete which makes the resistance measurement much more time efficient in the long run.

\subsection{Graphical User Interface (GUI)}

\begin{figure} [H]
    \centering
    \includegraphics[scale=1.2]{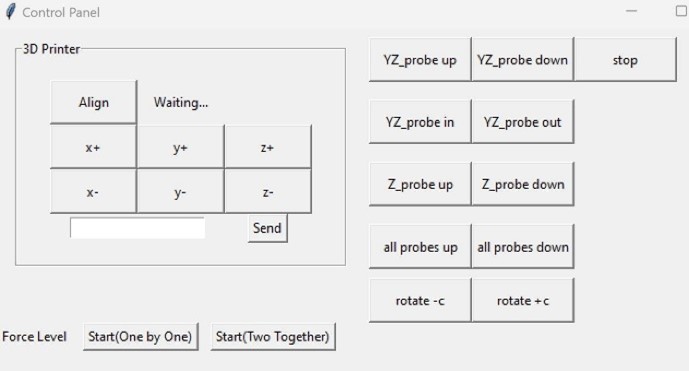}
    \caption{This figure shows the GUI control panel developed using Python Tkinter. It has three components. The part inside the “3D Printer" frame is used to control the 3D printer stepper motors, as well as the auto-alignment. The  part on the right of the “3D Printer" frame is used to control the motorised linear stages. The top four rows are used to control the YZ and Z probes respectively. The bottom button named “rotate” are used to rotate the rotation stage in clockwise “+c” and anticlockwise “-c” directions respectively to change the angle between the JJ pads and the probes. The part below the “3D Printer" frame shows the force level exerted by the probes on the qubit, and the buttons are used to start the auto-lowering of the probes which will be explained in section 3.2.}
    \label{fig:gui}
\end{figure}

A graphical user interface (GUI) based on Python Tkinter is designed to facilitate the supervision and control of the system. The GUI design is shown in Fig \ref{fig:gui}. The buttons in the “3D Printer" frame are used to control the 3D printer stepper motors. The “Align” button is used to start the auto-alignment which will be further explained in section 3.1. The x+, x-, y+, y-, z+, and z- buttons are used to move both the microscope camera and probes in positive and negative x, y, and z axis respectively since they replace the extruder. Each click makes a movement of 0.1 mm in the respective direction. The positive and negative x and y directions are with respect to the camera view. Horizontally right is the positive x direction and vertically up is the positive y direction. However, this positive y direction is opposite to the actual positive y direction in the actual 3D printer since the camera image is vertically mirrored by the microscope camera. The entry box and “Send” button are used to send customised G-code commands to the 3D printer if the operator wants the printer to have other movements.

The buttons on the right of the “3D Printer" frame are used to control the stepper motors connected to the Arduino Uno microcontrollers. The YZ probe is controlled by buttons labeled “YZ probe” while the Z probe is controlled by buttons labeled “Z probe”. The simultaneous control of the two probes is via buttons labeled “all probes”. The “up” and “down” indicates the direction of the probe movement. The “rotate” is used to change the angle of the rotation stage placed under the mounting platform so that the JJ pads can align with the probes. “+c” means the rotation stage will rotate in the clockwise direction in the camera view while “-c” means anti-clockwise direction. All motions of the Arduino motors can be stopped instantly by the “stop” button.

The buttons below the “3D Printer" frame are used for the auto-lowering. There are two start buttons. They are for auto-lowering of probes one by one and two probes together respectively. They will be explained further in section 3.2. The force level detected by the force sensor is shown at the position labeled “Force Level” for auto-lowering and human supervision purposes. 

\begin{figure}
    \centering
    \includegraphics[scale=1]{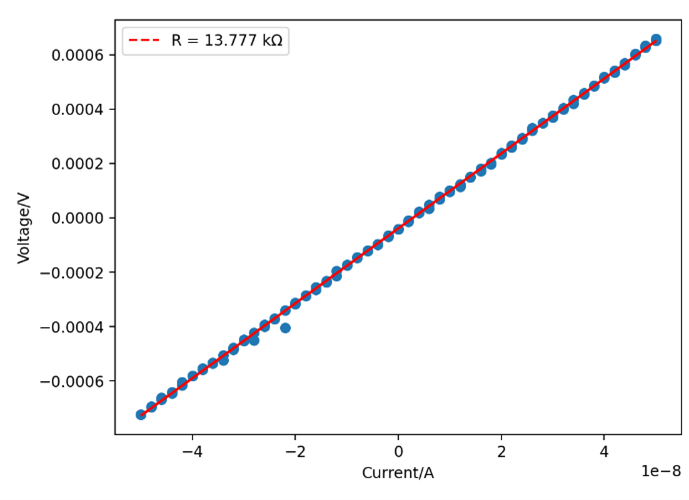}
    \caption{This Voltage/V against Current/A graph is used to calculate the resistance of the JJ which is the gradient of the linear fit line of the data points. This is an automated resistance measurement program that sets limits for both maximum voltage and current supply to protect the JJ \cite{Soe}.}
    \label{fig:measurement}
\end{figure}

After the probes touch the JJ pads properly, the resistance measurement program controlling a Keysight device is ready to be launched \cite{Soe}. It limits the maximum voltage and current to prevent damaging the JJ during the measurement. The resistance measured is the gradient value of a voltage-current graph shown in Fig \ref{fig:measurement}.

\section{Automated Processes}
\subsection{Auto-alignment}

\begin{figure} [H]
    \centering
    \includegraphics[scale=0.4]{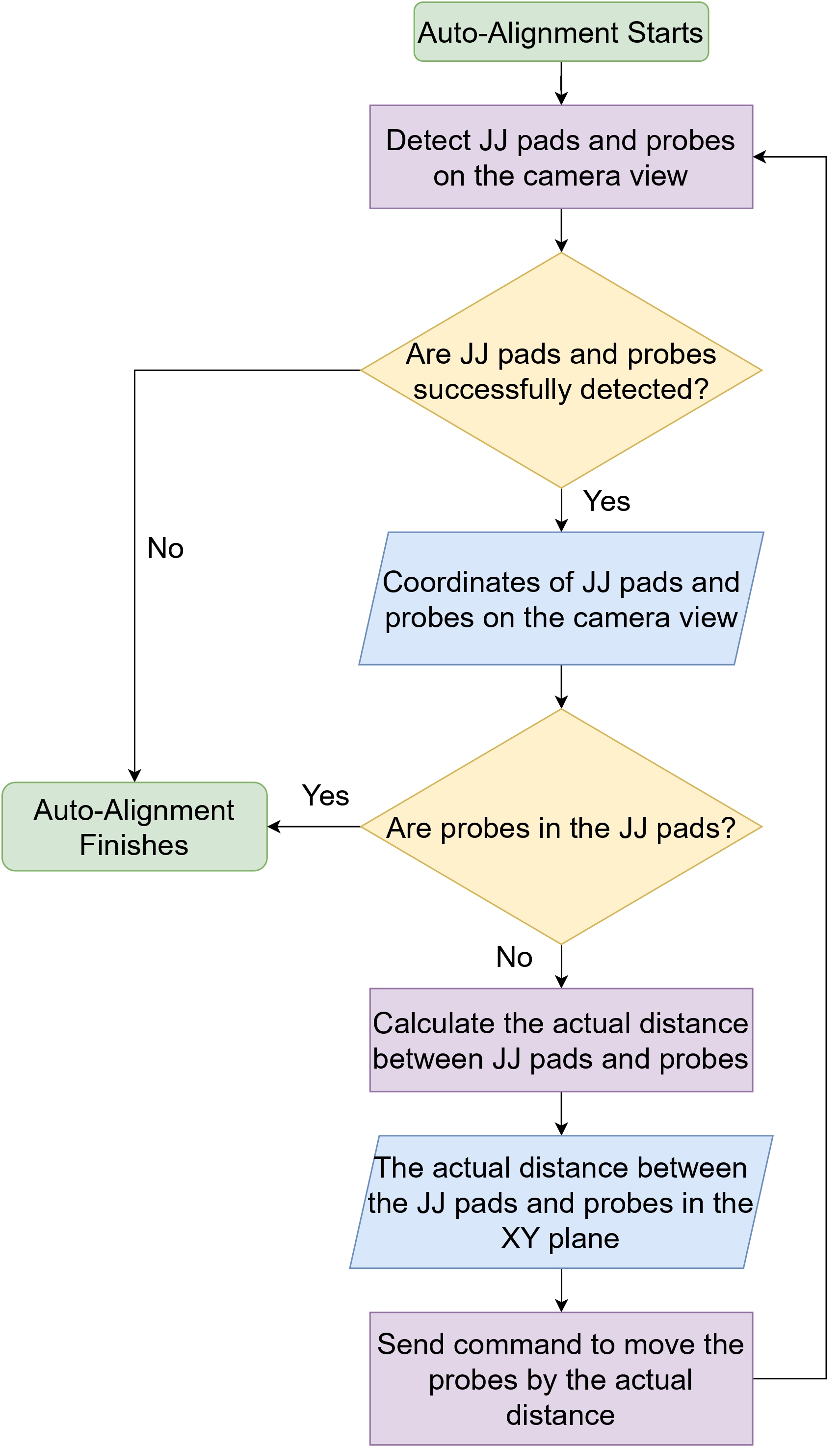}
    \caption{Flowchart Diagram of the Auto-alignment Process}
    \label{fig:auto-alignment}
\end{figure}

Before starting the alignment, the distance between the two probes, and the angle between the JJ pads and the probes are adjusted accordingly so that the probes can be placed on the JJ pads. 

The first automation process is auto-alignment. A Tensorflow object detection model is trained and loaded beforehand to allow the system to detect probes and JJ pads. A flowchart of this process is shown in Figure \ref{fig:auto-alignment}.

The object detection program can read out the coordinates of each side of every box on the camera view. This allows the system to calculate the actual distance between JJ pads and the tips of the probes. To calculate this actual distance, an additional correction factor is required to calibrate the coordinates on the camera view to the actual distance. This correction factor is the scale of the difference between the coordinates of two fixed points and the actual distance, and it can be experimentally determined. Since the Z-coordinate of the camera is never changed to keep its sharp focus on the qubit, the correction factor is constant. The actual distance between these two points is obtained by multiplying the difference between the two coordinates by the correction factor. With this actual distance, the system will send a G-code command to the 3D printer to move the probes by that actual distance to align the probes with the JJ pads. However, due to the occasional belt slipping when changing the X and Y coordinates of the probes, the probes may not move the same distance as what the system commands. Thus, the confirmation of the proper alignment is necessary. If the probes are not properly aligned, the system will repeat the above-mentioned procedure until the coordinates of the probes and JJ pads coincide.

\subsection{Auto-lowering}

\begin{figure} [H]
    \centering
    \includegraphics[width=\textwidth]{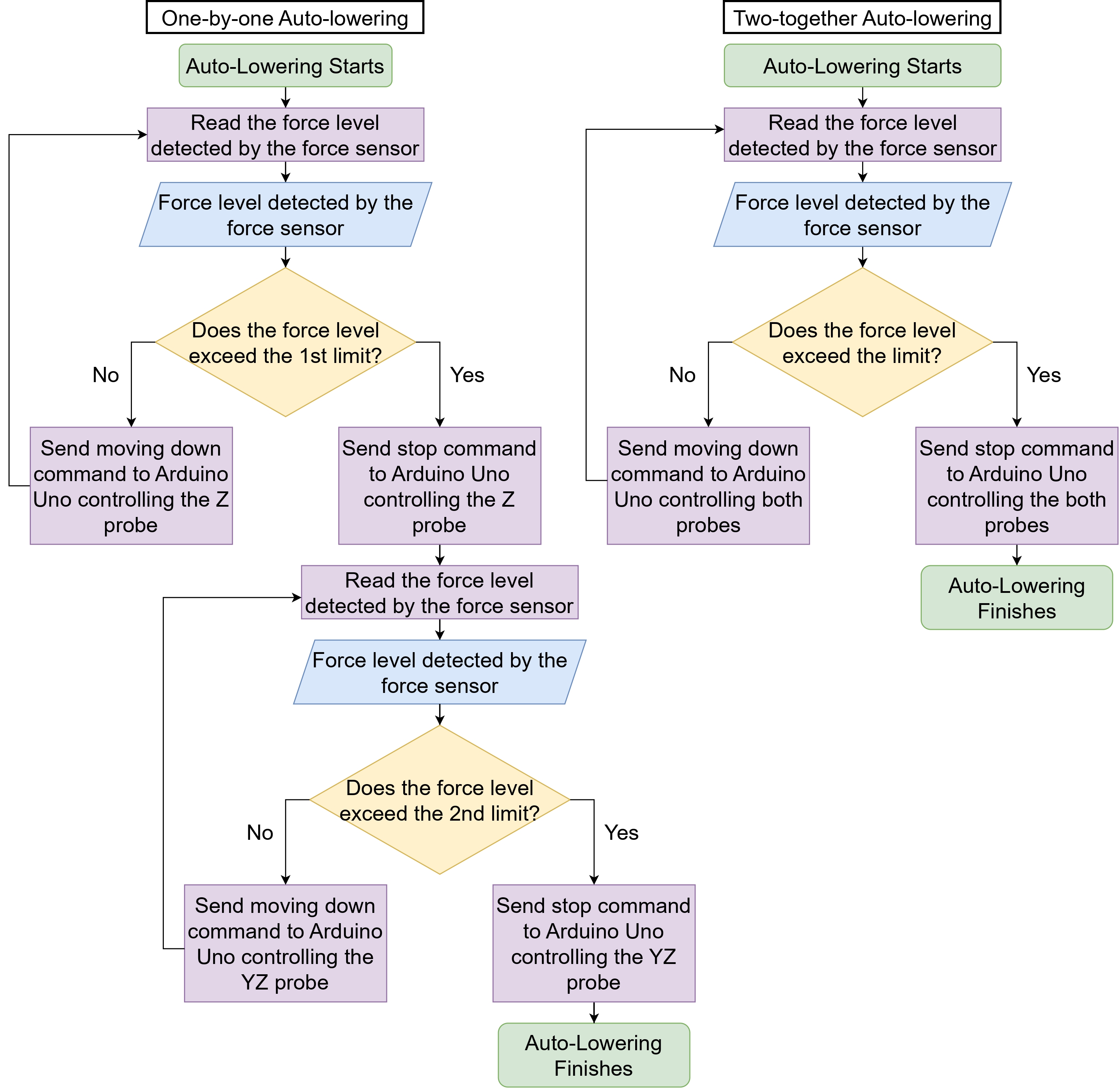}
    \caption{Left: Flowchart Diagram of the Auto-lowering of One Probe Each Time. Right: Flowchart Diagram of the Auto-lowering of Two Probes Together.}
    \label{fig:auto-lowering}
\end{figure}

After the proper alignment is finished, the system will start the auto-lowering process. The flowchart of this process is shown in Figure \ref{fig:auto-lowering}.

Since two probes may not have the same Z-coordinate at the beginning of the lowering, in the first auto-lowering, the probes are needed to be lowered one by one to align their Z-coordinates. This is to prevent one probe from exerting too much force onto the qubit and damaging the JJ if one probe reaches the qubit earlier than the other probe. All pre-set limits of the force are experimentally determined. 

The Z probe will lower first, and it will stop when the force level reading exceeds the first pre-set limit which is 0.05 N. After the Z probe is lowered, the YZ probe will start to lower and if the force level exceeds the second pre-set limit, which is 0.1 N, the YZ probe will stop lowering.

After this one-by-one lowering, the probes should have the same z-coordinate. Hence, the two probes lowering can then be safely carried out for the subsequent measurements to save the lowering time.
When the two-probes auto-lowering starts, the system will start to read the force level detected by the force sensor. Both probes are lowered at the same time. If the force level exceeds 0.1 N, the lowering will stop.

\section{Results}
For the measurement at the manual probe station, the average time taken for the alignment is 10-20 seconds depending on the distance between the probes and the JJ pads. The average time taken for the lowering is about 10-15 seconds.

On average, the time taken for auto-alignment is about 1 second. The precision of the auto-alignment relies on the precision of the pre-trained Tensorflow object detection model. The alignment can also be done with motor control if the auto-alignment is not precise. It would usually take 5-10 seconds to do the manual alignment. The average time taken for the lowering of two probes together is about 6-8 seconds, and it takes 12-16 seconds if the probes are lowered one by one.

The time taken for the resistance measurement via the Keysight device is 20 seconds.

As a result, the total time taken for automated measurement when the probes are auto-aligned and lowered two-together is 27-29 seconds. At the manual probe station, the total time taken is 40-55 seconds. Hence, the automated resistance measurement system can save time by 28-51\% compared to the manual probe station.

\section{Further Work}
First, the system can be further programmed to automatically measure the resistance of multiple JJs one by one. It can be done by using 3D-printed slots as the mounting platform of the qubits. Thus, the approximate coordinates of each qubit on the 3D printer frame are fixed. After measuring the resistance of one qubit, the probes can automatically move to the next qubit for subsequent measurements. If the model is trained to be sufficiently precise, the whole measurement process of multiple qubits can be done without human supervision.

Second, a high-power infrared laser can be added to the system. This high-power laser serves as a heating source to anneal the JJ to increase its normal state resistance, thus optimizing the junctions’ frequency \cite{Hertzberg2021, rosenblatt2019laser}.

\section{Conclusion}
To conclude, this paper presents the development of an automated superconducting qubit characterisation process based on a modified 3D printer with multiple Arduino Uno microcontrollers and motorised linear stages. The automation is achieved via auto-alignment and auto-lowering. It saves the measurement time by 28-51\% compared to the manual probe station and has the potential to be fully unsupervised. It is also much more affordable with only 800 SGD than other comparable commercial solutions. 

\bibliographystyle{plain}
\bibliography{reference}

\end{document}